\documentclass[twocolumn,aps,prb,showpacs]{revtex4}
\usepackage{epsf}
\usepackage{bm}

\begin{document}

\title {
         The electronic structures, the equilibrium geometries
         and finite temperature properties of Na$_n$ ($n$=39-55)
       }

\author {Shahab Zorriasatein$^{1,2}$, Mal-Soon Lee$^1$, and D. G. Kanhere$^1$}

\affiliation {
    $^1$ Department of Physics, and
         Center for Modeling and Simulation, 
         University of Pune, 
         Ganeshkhind, 
         Pune--411 007, 
         India                                 \\
    $^2$ Department of Physics, 
         Islamic Azad University, 
         Tehran south branch, 
         Tehran, Iran
             }

\begin{abstract}
Density-functional theory has been applied to investigate systematics of  
sodium clusters Na$_n$ in the size range of $n$= 39-55. A clear 
evolutionary trend in the growth of their ground-state geometries emerges.
The clusters at the beginning of the series ($n$=39-43) are symmetric and have 
partial icosahedral (two-shell) structure. The growth then goes through a 
series  of disordered clusters ($n$=44-52) where the icosahedral core is lost.
However, for $n\ge$53 a three shell icosahedral structure emerges. This change
in the nature of the geometry is abrupt. In addition, density-functional 
molecular dynamics has been used  to calculate the specific heat curves for
the representative sizes  $n$= 43, 45, 48 and 52. These results along with 
already available  thermodynamic calculations for $n$= 40, 50, and  55 enable us to 
carry out  a detailed comparison of the heat capacity curves with their 
respective geometries  for the entire series. Our results clearly bring out 
strong correlation between  the evolution of the geometries and the nature of 
the shape of the heat  capacities. The results also firmly establish the 
size-sensitive nature of the heat  capacities in sodium clusters.
 
\end{abstract}


\maketitle

\section{Introduction}
Physics and chemistry of clusters are very active areas of research especially
because of  the emergence of nano science and nano technology.~\cite{nano}
Although major  efforts have been spent into ground-state investigations,
finite temperature  properties are turning out to be very interesting. 
Such investigations are challenging, both experimentally as well as 
theoretically. One of the first  detailed measurements providing much impetus
for theoretical work was on free sodium clusters by Haberland and 
co-workers.~\cite{Haberland} These measurements reported the  melting temperatures 
(T$_m$) of sodium clusters in the size range between 55 and 350 and remained 
unexplained for almost about a decade.

The main puzzle was  related to the irregular  behavior of  the melting
temperature and the absence of any correlation between the peaks and the
magic numbers either  geometric or electronic.
A good deal of simulation works has been carried out to explain the sodium
data, most of the early work being with classical inter atomic 
potentials.~\cite{Calvo, Li} It turned out that none of these could obtain 
qualitative and  quantitative agreement with the experimental data. 
Thus, it needed an {\it ab initio} density-functional method to achieve this. 
Indeed, much insight and excellent quantitative  agreement has been
obtained by density-functional molecular dynamics (DFMD) 
simulations.~\cite{Manninan, Chacko, Lee-small-sodium,Lee-58, Aguado, Aguado30}

Recently a very different aspect of finite temperature behavior has been
brought out by the experimental and later by theoretical work on gallium
clusters.~\cite{Jarrold-Ga, Kavita} The  experimental reports of Breaux and
co-workers~\cite{Jarrold-Ga} showed that in the size range  of $N$=30 to 55,
free clusters of gallium melt much above their bulk melting  temperature.
Interestingly, their experiment also showed that the nature of heat capacity 
is  size sensitive.
In fact, addition of even one atom changes the shape of the heat capacity
dramatically, e.g for Ga$_{30}$ and Ga$_{31}$.  
A similar experimental observation has been reported for aluminum clusters
in the size range $n$=49-62.~\cite{Jarrold-Al} Such a size sensitive 
behavior has also been observed in DFMD simulation of Au$_{19}$ and 
Au$_{20}$.~\cite{Au19-20}
A detailed analysis of the ground-state geometries of these clusters
brought out the  role of  order and disorder in their geometries on the
shape of the melting curve. A disordered system is shown to display a
continuous melting transition leading to a very broad heat capacity curve.
However, the effect is subtle and description of order and disorder needs careful
qualifications. 
 
In spite of  substantial experimental works on sodium clusters over a period 
of 10 years and or so, there is no firm and systematic evidence of size
sensitivity.
This is mainly due to the fact that the reported experimental
data~\cite{Haberland} is for the size range of $n$=55-350 at discrete sizes.
It is necessary to investigate  the effect of addition of few atoms in a 
continuous manner in appropriate size range. However, it is not clear whether 
larger clusters having sizes of $n>$100 will also show this effect.

An extensive {\it ab initio} study on the structural properties of small Na clusters
up to $N$=20 has been reported by R\"othlisberger and Andreoni.~\cite{Rothlisberger}  
The study reveals that pentagonal motifs dominate the structures above $N$=7. 
As expected most of the atoms in these clusters lie on the surface and 
a discernible core develops after about $N$=15-16. 
The shapes after this sizes show  signature of icosahedral structures.

The finite temperature behavior of sodium clusters in the size range of
$n$=8 to 55 has been reported.~\cite{Lee-small-sodium} The study reveals that
it is not easy to discern any melting peaks below $n$=25. However, the 
simulation data available at rather coarse sizes above $n$=40 already 
shows size-sensitive feature. 
In addition to this feature, our recent investigation~\cite{Lee-58} on Na$_{57}$ and 
Na$_{58}$ does bring out the role of geometry and electronic structure on the melting.
Nevertheless, in order to draw  definitive conclusions it is necessary to 
investigate  the effect of addition of a few atoms in a  continuous manner 
in appropriate size range.
Therefore, we have chosen the size range of $n$=39-55 and have carried out 
detailed density-functional investigations.
The purpose of the present work is two fold. First, to obtain reliable 
equilibrium geometries for all the clusters in the size range of $n$=39-55
and to discern evolutionary trends. We note that Na$_{40}$ has a symmetric
partially  icosahedron  core and Na$_{55}$ is a complete  icosahedron.
Thus it is of considerable interest to examine the growth pattern from $n$=39
to $n$=55.
The second purpose is to seek correlation between the nature of the ground-state 
and the evolutionary trends observed in the nature of their specific 
heats. Towards this end we have carried out extensive finite temperature 
simulations  on representative clusters of size $n$= 43, 45, 48, and 52.
Together with the already published results, this gives us access to  specific 
heats for Na$_{40}$, Na$_{43}$, Na$_{45}$, Na$_{48}$, Na$_{50}$, Na$_{52}$
and  Na$_{55}$, a reasonable representation across the series under 
investigation.
Finally, we note that all the DFMD simulations reported so far have yielded
excellent agreement with the experimental data~\cite{Aguado,Chacko,Lee-58} 
These reports demonstrate the reliability of density-functional molecular 
dynamics in describing the finite temperature properties.
  
The plan of the paper is as follows. 
In the next section (Sec.~\ref{sec:comp}) we note the computational details. 
Sec.~\ref{sec:geom} presents equilibrium geometries and their shape systematic.  
Sec.~\ref{sec:therm} presents the finite temperatures behavior of 
Na$_{43}$, Na$_{45}$, Na$_{48}$, Na$_{52}$ and finally we discuss the 
correlation between the ground states and nature of the specific heats for
all the available thermodynamics data. We conclude our discussion in 
Sec.~\ref{sec:concl}.

\section{Computational Details}\label{sec:comp}
We have carried out Born-Oppenheimer molecular dynamics (BOMD)
simulations~\cite{MD} using Vanderbilt's ultrasoft pseudopotentials~\cite{vanderbilt}  
within the local-density approximation (LDA), as implemented in the 
VASP package.~\cite{VASP}
We have optimized about 300 geometries for each of the sodium clusters in 
the size range between $n$=39 and $n$=55. The initial configuration for the optimization
of each cluster were obtained by  carrying out a constant temperature
dynamics simulation of 60~ps each at various temperatures between 300 to 400~K.
For many of the geometries we have also employed basin hopping~\cite{Basin}
and genetic~\cite{genetic1,genetic2} algorithms using Gupta potential~\cite{Li} 
for generating initial guesses.
Then we optimized these structures by using the {\it ab initio}
density-functional  method.~\cite{ab initio}
For computing the heat capacities, the iso-kinetic BOMD calculations were carried out at 
14  different  temperatures for each cluster of Na$_{43}$, Na$_{45}$, 
Na$_{48}$, Na$_{52}$  in  the range between 100~K and 460~K, each with the time  duration 
of  180~ps or more.
Thus, it results in the total simulation time of 2.5~ns per system. 
In order to get converged heat capacity curve especially in the region of co-existence,  
more  temperatures were  required with longer simulation times. We have discarded 
at least first 30~ps for each temperature for thermalization.
To analyze the thermodynamic properties,  we first calculate the ionic 
specific heat by using the Multiple Histogram (MH) technique.~\cite{Mh1, Mh2}
We extract the  classical ionic density of states ($\Omega (E)$) of 
the system, or equivalently  the classical ionic entropy, $S(E)=k_{B}\ln \Omega (E)$, 
following the MH technique. 
With $S(E)$ in hand, one can evaluate thermodynamic averages in a variety of ensembles.  
We focus in this work on the ionic specific heat. 
In the canonical ensemble, the specific heat is defined as
usual by $C(T)=\partial U(T)/\partial T$, where $U(T)=\int E\,p(E,T)\,dE$
is the average total energy. The probability of observing an
energy $E$ at a temperature $T$ is given by the Gibbs distribution
$p(E,T)=\Omega (E)\exp (-E/k_{B}T)/Z(T)$, with $Z(T)$ the normalizing
canonical partition function. 
We normalize the calculated canonical  specific heat by the zero-temperature
classical limit of the rotational  plus vibrational specific heat,
i.e., $C_{0}=(3N-9/2)k_{B}$.~\cite{Abhijat}

We have calculated a number of thermodynamic indicators such as  
root-mean-square bond length fluctuations  ($\delta_{\rm rms}$), mean square
displacements (MSD) and radial distribution function (g(r)).
The $\delta_{\rm rms}$ is  defined as 

\begin{equation}
\delta _{{\rm rms}}=\frac{2}{N(N-1)}\sum_{i>j}\frac{(\langle
r_{ij}^{2}\rangle _{t}-\langle r_{ij}\rangle _{t}^{2})^{1/2}}{\langle
r_{ij}\rangle _{t}},  \label{eqn:delta}
\end{equation}

where $N$ is the number of atoms in the system, $r_{ij}$ is the distance
between atoms $i$ and $j$, and $\langle \ldots \rangle _{t}$ denotes a
time average over the entire trajectory.
MSDs for individual atoms is another traditional parameter
used for determining phase transition and is defined as,

\begin{equation}
\langle {\bf r}_{I}^{2}(t)\rangle =\frac{1}{M}
\sum_{m=1}^{M}
\left[ {\bf R}_{I}(t_{0m}+t)-{\bf R}_{I}(t_{0m})\right]^{2}
\label{eqn:msq}
\end{equation}

where {\bf R}$_I$ is the position of the  $I$$^{th}$ atom and we average over M 
different time origins t$_{0m}$ spanning over the entire trajectory.
The interval between the consecutive t$_{0m}$ for the average was taken to be
about 1.5~ps. The MSDs of a cluster indicate the displacement of atoms in the
cluster as a function of time.
The g(r) is defined as the average number of atoms within the region $r$ and
$r+dr$.

We have also calculated the shape deformation parameter ($\varepsilon_{def}$),
to analyze the shape of the ground state for all the clusters.
The $\varepsilon_{def}$ is defined as,

\begin{equation}
\varepsilon_{def} = \frac{2Q_{x}}{Q_y+Q_z}, 
\label{eqn:epspro}
\end{equation}

where $Q_x \geq Q_y \geq Q_z$ are the eigenvalues,  in descending
order,  of the quadrupole tensor

\begin {equation}
        Q_{ij} =  {\sum_{I}  R_{Ii}\, R_{Ij} }.
\end {equation}

Here $i$ and $j$ run from 1 to 3,  $I$ runs over the number of ions,
and $R_{Ii}$ is the i$^{th}$ coordinate of ion $I$ relative to the
center of mass (COM) of the cluster. A spherical system ($Q_{x}=Q_{y}=Q_{z}$)
has $\varepsilon_{def}$=1 and larger values of $\varepsilon_{def}$ indicates 
deviation of the shape of the cluster from sphericity.

\begin{figure}
  \epsfxsize=0.4\textwidth
  \centerline{\epsfbox{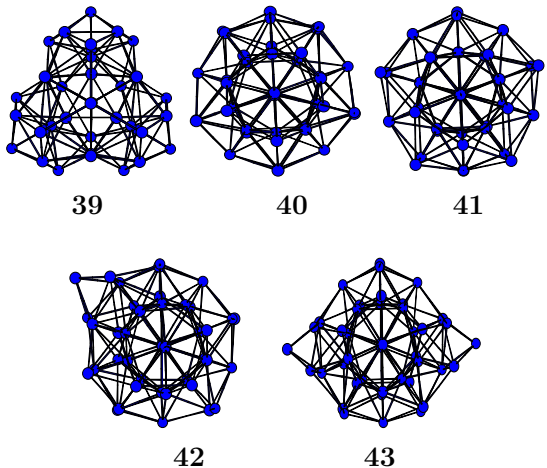}}
  \caption{\label{fig1}
  The ground-state geometries of Na$_{n}$ ($n$=39-43).
  }
\end{figure}
\begin{figure}
  \epsfxsize=0.4\textwidth
  \centerline{\epsfbox{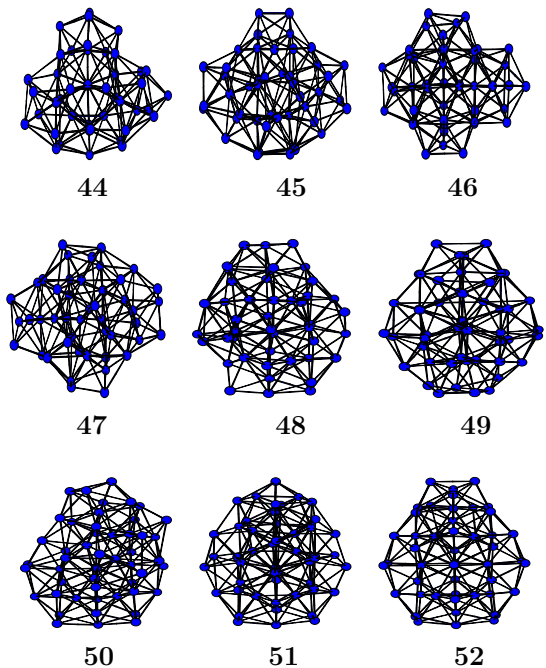}}
  \caption{\label{fig2}
  The ground-state geometries of Na$_{n}$ ($n$=44-52).
  }
\end{figure}
\begin{figure}
  \epsfxsize=0.4\textwidth
  \centerline{\epsfbox{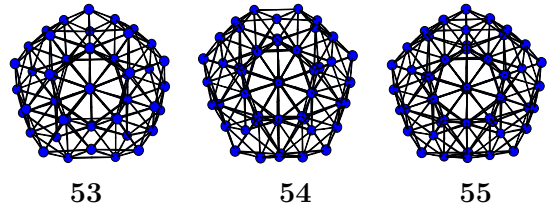}}
  \caption{\label{fig3}
  The ground-state geometries of Na$_{n}$ ($n$=53-55).
  }
\end{figure}

\section {Geometries}\label{sec:geom} 

\begin{figure}
  \epsfxsize=0.35\textwidth
  \centerline{\epsfbox{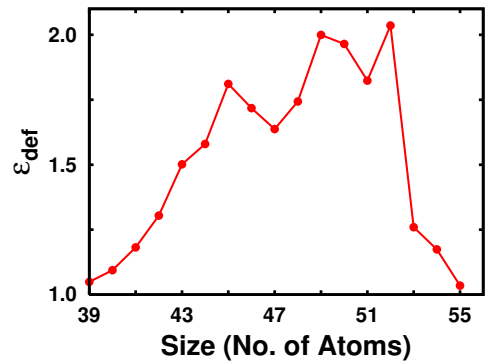}}
  \caption{\label{fig4}
  The shape deformation parameter for Na$_{n}$ ($n$=39-55) as a function of cluster
  size.
  }
\end{figure}

\begin{figure}
  \epsfxsize=0.45\textwidth
  \centerline{\epsfbox{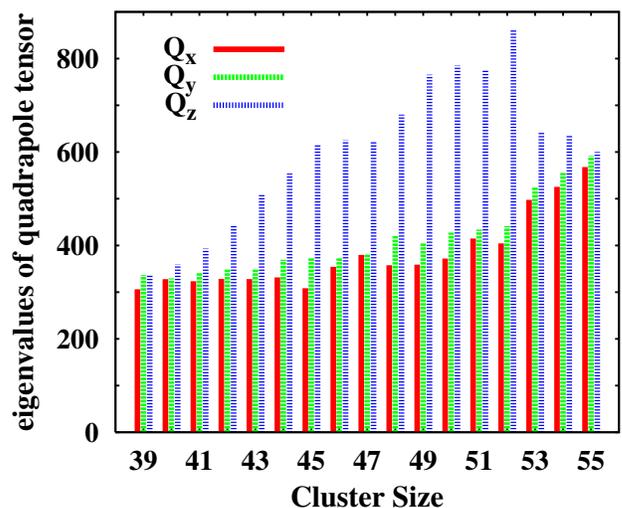}}
  \caption{\label{fig5}
  The eigenvalues of the quadrupole tensor for Na$_{n}$ ($n$=39-55) as a
  function of cluster size.
  }
\end{figure}

The lowest energy geometries of sodium clusters (Na$_{n}$ $n$=39-55) are shown
in Fig.\ \ref{fig1} ($n$=39-43), Fig.\ \ref{fig2} ($n$=44-52), and Fig.\ \ref{fig3}
($n$=53-55).
We have also plotted the shape deformation parameter $\varepsilon_{def}$ and 
the  eigenvalues of quadrupole tensor for the ground state geometries of these 
clusters in Fig.\ \ref{fig4} and Fig.\ \ref{fig5}, respectively. 

\begin{figure*}
  \epsfxsize=0.63\textwidth
  \centerline{\epsfbox{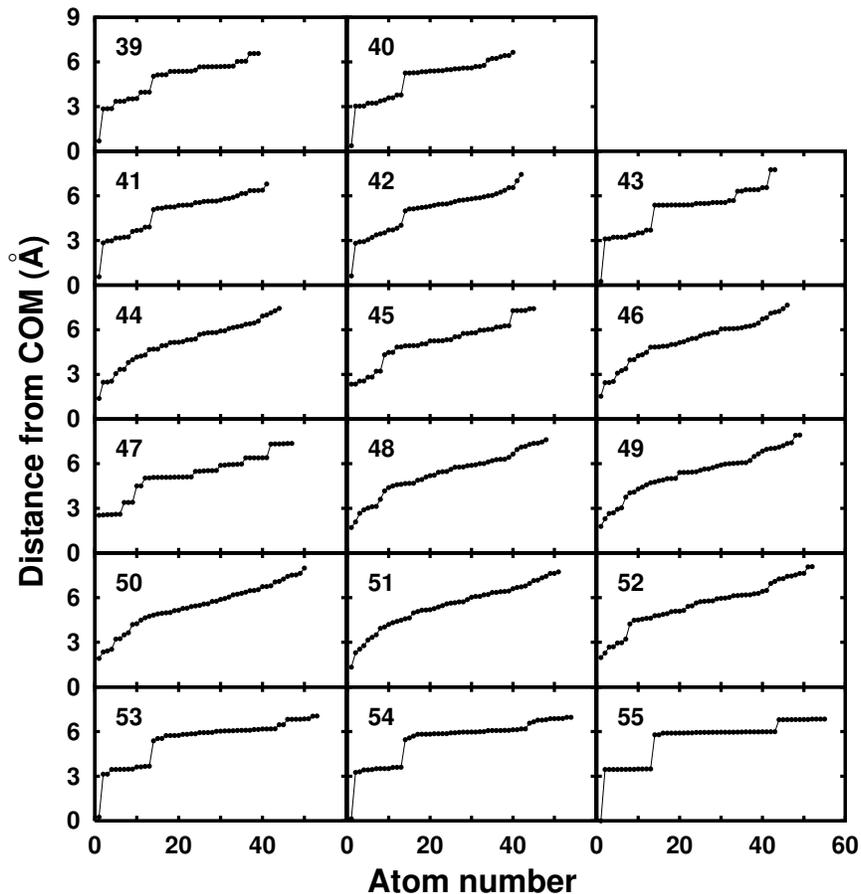}}
  \caption{\label{fig6}
  The distance from the center of mass for each of the atoms ordered
  in the increasing fashion for  Na$_n$ ($n$=39-55). The formation of 
  the shells is evident from the sharp steps.
  }
\end{figure*}

It is convenient to divide these clusters into three groups. The clusters 
in the first group, shown in Fig.\ \ref{fig1} are nearly  spherical.
The ground-state geometry for  Na$_{39}$ is highly symmetric. This structure
has three identical units, each based on the icosahedral motive and these 
three units are arranged as shown in the Fig.\ \ref{fig1}.
It is interesting to note that an addition of an extra atom changes the
structure dramatically. It can be seen that the geometry of Na$_{40}$ is
based on icosahedral structure with missing 12 corner atoms and (111) facet 
as reported by Rytk{\"o}nen {\it et al} .~\cite{Manninan} 
An extra atom added to this structure is accommodated near the surface 
and deforms the structure slightly. In addition to the deformation the 
distance  between the two  shells is reduced by 0.3~\AA\ as compared to
that in Na$_{40}$.
A single atom added to Na$_{41}$ is not accommodated in the structure, 
instead  it caps  the surface.
However, the low-lying geometries for Na$_{42}$ have a spherical shape 
without any cap (figure not shown). The lowest-energy structure of Na$_{43}$ shows
two caps symmetrically placed on the  opposite side of Na$_{41}$, accompanied
by distortion of the icosahedral core.
The second group in Fig.\ \ref{fig2} consisting of the clusters with $n$=44-52
shows substantial distortion of the icosahedral core and even  loss of this
core structure. These clusters essentially represent the transition region
from the two shell icosahedron, Na$_{40}$ to three shell complete icosahedron,
Na$_{55}$.
It can be seen that by adding one atom to  Na$_{43}$ the two shell core is
destroyed. The growth from $n$=44 to $n$=52 shows successive stages of capping
followed  by rearrangement in the inner core. 
There is a dramatic change in the structure as soon as we add one more 
atom  to Na$_{52}$. All the atoms rearrange to form an icosahedral structure
as seen in  Na$_{53}$.
Thus, the clusters in the last group namely, Na$_{53}$ and Na$_{54}$ differ 
from a perfect icosahedron of Na$_{55}$ by an absence of two and one atom(s), 
respectively (Fig.\ \ref{fig3}).

The nature of the changes in the shape of the clusters during the growth
can be seen in Fig.\ \ref{fig4} and Fig.\ \ref{fig5}.
The shape deformation parameter ($\varepsilon_{def}$) increases to a value 
about 2 up to Na$_{52}$ with slightly higher values for 
Na$_{45}$ and Na$_{49}$ (Fig.\ \ref{fig4}). 
However, this value drops suddenly for Na$_{53}$.
It is interesting to examine the three eigenvalues of quadrupole tensor
($Q_x,~Q_y,~Q_z$) shown in Fig.\ \ref{fig5}.
It can be seen that two of the eigenvalues are nearly same up to  Na$_{52}$
while the third one continuously grows and indicates that the growth
dominantly takes place along one of the directions. A prolate configuration
has $Q_z\gg Q_x\approx Q_y$. Thus, the majority of the clusters in
the second group are prolate.  
The formation, the destruction and reformation of the shell structure is 
clearly seen in Fig.\ \ref{fig6}.
In this figure we have plotted the distance of each atom from
the center of mass arranged in the increasing fashion.
Clearly small rearrangement of atoms yields a change in the structure 
from  Na$_{39}$ to Na$_{40}$. The two shell structures are observed till
Na$_{43}$. The formation of the shell structure  is reflected in the 
formation of the sharp steps  in the graph.
As the size increases from Na$_{43}$ to Na$_{44}$ the shell
structure is destroyed and seen again at Na$_{53}$. Thus, three shells begin to
emerge at Na$_{53}$.

\section {Thermodynamics}\label{sec:therm}
We have calculated  the ionic heat capacity and indicators like mean square 
displacements (MSD) and root-mean-square bond length fluctuations 
($\delta_{\rm rms}$) for four of the representative clusters in the 
investigated series which are Na$_{43}$, Na$_{45}$, Na$_{48}$, and Na$_{52}$.

\begin{figure}
  \epsfxsize=0.34\textwidth
  \centerline{\epsfbox{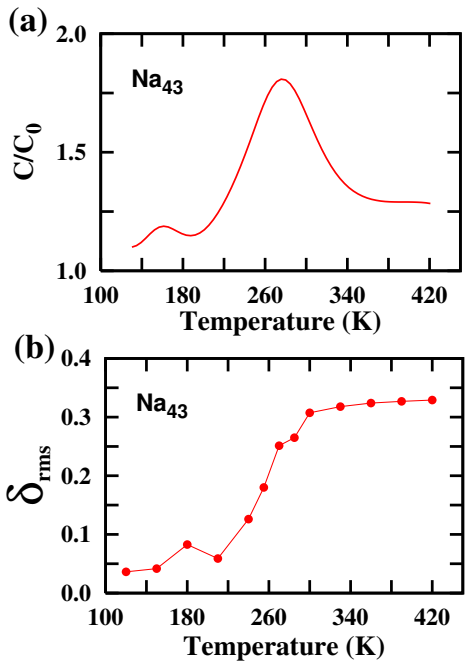}}
  \caption{\label{fig7}
  (a) The normalized heat capacity and (b) the $\delta_{\rm rms}$ for Na$_{43}$. 
  The peak in heat capacity curve is at 270~K.
  }
\end{figure}
\begin{figure}
  \epsfxsize=0.32\textwidth
  \centerline{\epsfbox{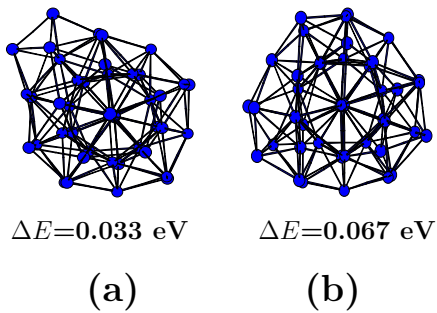}}
  \caption{\label{fig8}
  Two low lying isomers of Na$_{43}$.
  $\Delta$E represents the energy difference with respect to the ground-state.
  }
\end{figure}

\begin{figure}
  \epsfxsize=0.32\textwidth
  \centerline{\epsfbox{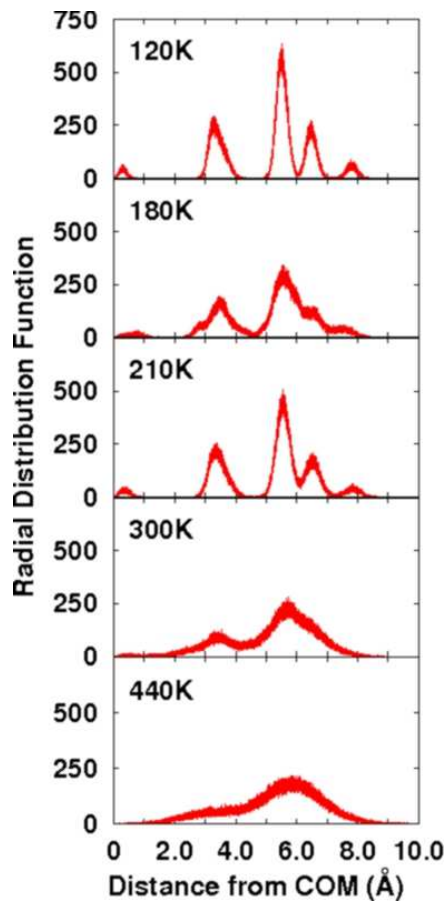}}
  \caption{\label{fig9}
  The radial distribution function calculated for Na$_{43}$ at five different
  temperatures.
  }
\end{figure}

We note that the thermodynamics of Na$_{40}$~\cite{Manninan,Lee-small-sodium},
Na$_{50}$~\cite{Lee-small-sodium} and Na$_{55}$~\cite{Aguado, Chacko} has 
already been reported.
Thus it is possible to examine and analyze the systematic variations in the
melting characteristic and correlate them with the equilibrium geometries 
across the entire range of sizes from 40 to 55. 
The heat capacity and $\delta_{\rm rms}$ for  Na$_{43}$ are shown in 
Figs.\ \ref{fig7}(a) and \ref{fig7}(b), respectively. We also show typical
low energy geometries (isomers) for Na$_{43}$ in Fig.\ \ref{fig8}. The first
isomer shown in  Fig.\ \ref{fig8}(a) has two atoms closest to each other 
capping the surface and second isomer shows a distorted shape and no caps.
The heat capacity shows a weak peak  around 160~K while the main peak 
occurs at 270~K. An examination of the motion of ionic trajectory seen as 
a movie indicates that isomerization (Fig.\ \ref{fig8}(a)) is responsible 
for the weak peak.
\begin{figure}
  \epsfxsize=0.34\textwidth
  \centerline{\epsfbox{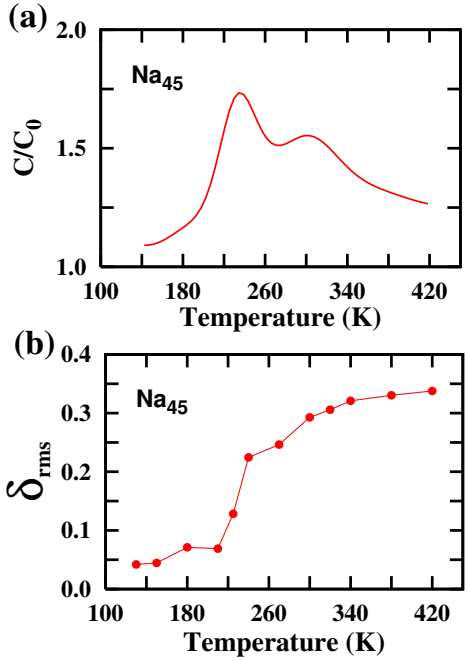}}
  \caption{\label{fig10}
 (a) The normalized heat capacity and (b) the $\delta_{\rm rms}$ for Na$_{45}$.
 The heat capacity curve shows a two stages melting process.}
\end{figure}

\begin{figure}
  \epsfxsize=0.32\textwidth
  \centerline{\epsfbox{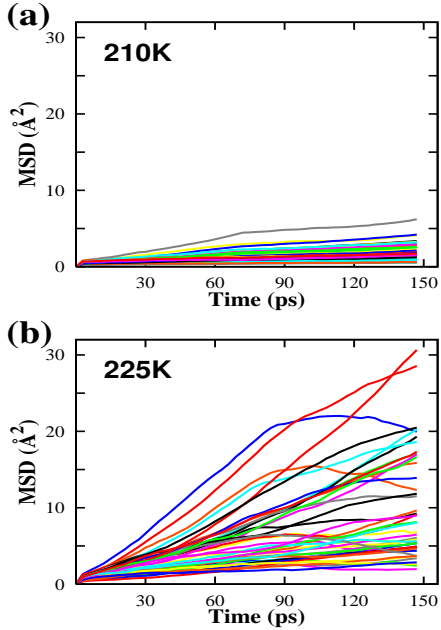}}
  \caption{\label{fig11}
  The MSDs of individual atoms calculated for Na$_{45}$ (a) at 210~K and (b) 
  at 225~K over the last 150~ps.}
\end{figure}

It is interesting to observe the changes of radial distribution function (RDF)
as a function of temperature which is shown in  Fig.\ \ref{fig9}. At low 
temperatures the shell structure is clearly evident. The pattern seen at 180~K
and 210~K are mainly due to the fluctuation of the cluster between the
ground-state and low-lying states. At 300~K and above the RDF shows the typical
melting behavior of a cluster. The $\delta_{\rm rms}$ in Fig.\ \ref{fig7}(b)
shows  the effect of isomerization around 160~K. It can be seen that the melting
region is of the order of 60~K.

\begin{figure}
  \epsfxsize=0.34\textwidth
  \centerline{\epsfbox{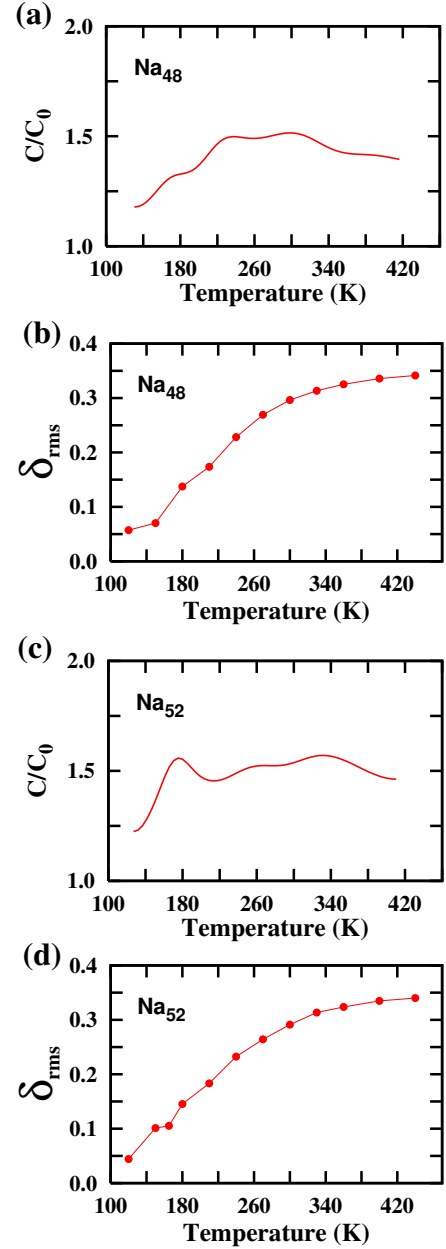}}
  \caption{\label{fig12}
  (a) The normalized heat capacity and (b) the $\delta_{\rm rms}$ for Na$_{48}$,  
  (c) The normalized heat capacity and (d) the $\delta_{\rm rms}$ for Na$_{52}$. 
  }
\end{figure}
\begin{figure}
  \epsfxsize=0.35\textwidth
  \centerline{\epsfbox{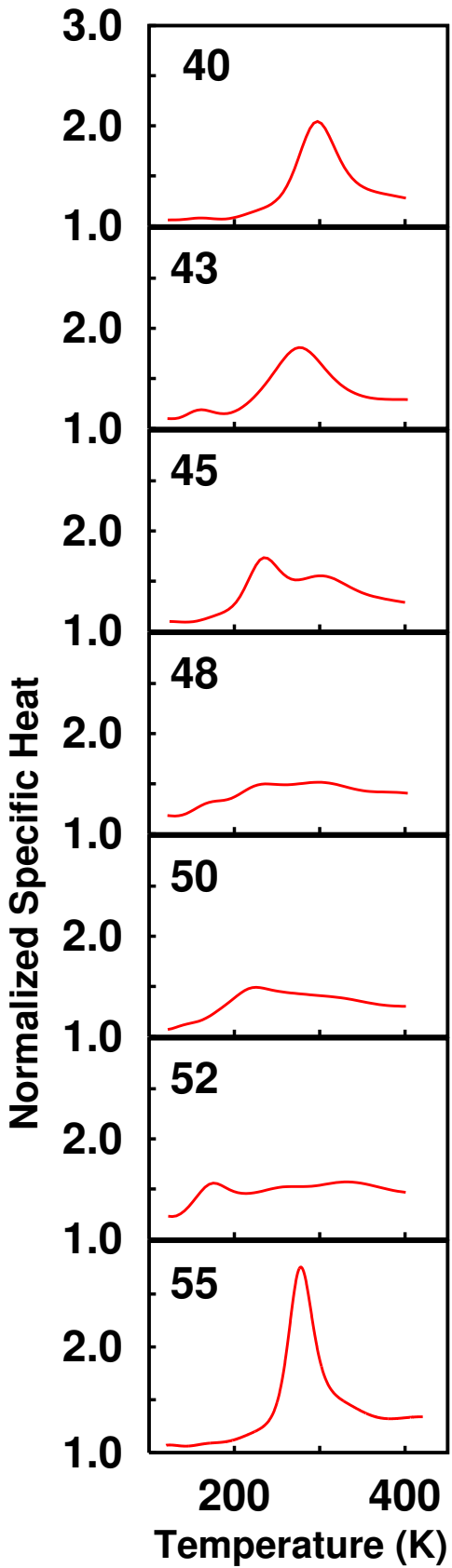}}
  \caption{\label{fig13}
  The normalized heat capacity as a function of temperature for Na$_n$, $n$=40,
  43, 45, 48, 50, 52, and 55.
  }
\end{figure}

Let us recall that Na$_{45}$, Na$_{48}$ and Na$_{52}$ belong to the second 
class of disordered clusters. Among these Na$_{45}$ shows some signature of 
partial shells.
The heat capacity and $\delta_{\rm rms}$ for  Na$_{45}$ are shown in 
Figs.\ \ref{fig10}(a) and \ref{fig10}(b), respectively.
The heat capacity of Na$_{45}$ shows a first peak around 230~K and 
a second peak around 300~K. We also show the MSDs of individual atoms at 210 
and 225~K in Figs.\ \ref{fig11}(a) and \ref{fig11}(b), respectively. We have observed 
the ionic trajectories as a movie in the temperature at 225~K. It turns 
out that about one third of atoms in the cluster ``melt" at this temperature. 
This is brought out by the contrasting behavior of the MSDs as shown in 
Figs.\ \ref{fig11}(a) and \ref{fig11}(b).
Interestingly, all these 15 atoms are on the surface, indicating that surface melting 
takes place first. The peak around 225~K in the heat capacity is due to partial
melting of these surface atoms. We show the $\delta_{\rm rms}$ as a function of 
the temperature in Fig.\ \ref{fig10}(b). There is a sharp increases in 
$\delta_{\rm rms}$ around 210~K and a slow rise after 240~K consistent with the
behavior of the heat capacity. Thus in Na$_{45}$ melting takes place in two stages
over the range of 120~K.

The heat capacity and $\delta_{\rm rms}$ for Na$_{48}$ and Na$_{52}$ 
are shown in Fig.\ \ref{fig12}. The heat capacity of Na$_{48}$ (Fig.\ \ref{fig12}(a)) 
is very broad indicating almost continuous  phase change starting around 150~k.
Thus it is  difficult to identify a definite  melting temperature. 
A similar behavior is also seen in that of Na$_{52}$ except for the small peak
seen 180~K due to isomerization (Fig.\ \ref{fig12}(c)).
It is interesting to note that $\delta_{\rm rms}$ for both clusters also
show a gradual rise from about 150~K to 350~K indicating a continuous melting
transition as shown in Figs.\ \ref{fig12}(b) and \ref{fig12}(d).

The systematic evolution of heat capacities can be better appreciated by
examination of all the available data (calculated from density-functional 
simulation). In Fig.\ \ref{fig13} we show the specific heats for all the
available clusters between  Na$_{40}$ and Na$_{55}$.
The most symmetric cluster Na$_{55}$ shows the sharpest peak in the heat
capacity.
The heat capacity of Na$_{40}$ and Na$_{43}$ (partial icosahedral structures)
have well recognizable peaks which are broader than that of Na$_{55}$. 
The disordered phase of the growth is clearly reflected in the very broad
heat capacities seen around $n$=50.

\section{Summary and conclusion\label{sec:concl}}

The {\it ab initio} density-functional method has been applied to investigate
systematic evolutionary trends in the ground state geometries of the 
sodium clusters in the size range of $n$=39-55. The DFMD finite temperature
simulations have been carried out for representative clusters. A detailed 
comparison between the heat capacities and the geometries firmly establishes a
direct influence of the geometries on the shapes of the heat capacity curves.
The heat capacities show size sensitivity. The growth pattern shows a 
transition from ordered $\longrightarrow$ disordered $\longrightarrow$ ordered 
sequence. 
The corresponding heat capacities show a transition from peaked to a very
broad to peaked sequence.
It is seen that addition of a few atoms changes the shape of heat capacity 
very significantly.
We believe that the size sensitive feature seen our simulation is universal.
It may be noted that such a feature has been observed experimentally in the
case of gallium and aluminum and in the case of DFMD simulations for gold. We 
await the experimental measurements of the heat capacities on the sodium
clusters in these range showing the size sensitivity.

\section  {Acknowledgment}

We acknowledge partial assistance from the Indo-French center for Promotion
of Advance Research (India)/ Center Franco-Indian pour la promotion de la
Recherche Avanc\'ee (France) (IFC-PAR, project No; 3104-2). We would like to
thank Kavita Joshi and Sailaja Krishnamurty for a number of useful discussions.


\begin{thebibliography}{}

\bibitem{nano}
 P. G. Reinhard, E. Suraud,
{\it Introduction to Cluster Dynamics} (Wiley-VCH, Berlin, 2003)

\bibitem{Haberland}
  M. Schmidt, R. Kusche, B. von. Issendorff, and H. Haberland,
  Nature (London) {\bf 393}, 238 (1998); 
  M. Schmidt and H. Haberland, 
  C. R. Phys. {\bf 3}, 327 (2002); 
  H. Haberland, T. Hippler, J. Donges, O. Kostko, M. Schmidt and B. von Issendroff, 
  Phys. Rev. Lett. {\bf 94}, 035701 (2005)

\bibitem{Calvo}
  F. Calvo and F. Spiegelmann, 
  J. Chem. Phys. {\bf 112}, 2888 (2000)

\bibitem{Li}
  Y. Li, E. Blaisten-Barojas, and D. A. Papaconstantopoulos, 
  Phys. Rev. B {\bf 57}, 15519 (1998).

\bibitem{Manninan}
  A. Rytk{\"o}nen, H. H{\"a}kkinen, and M. Manninen,
  Phys. Rev. Lett. {\bf 80}, 3940  (1998).

\bibitem{Chacko}
  S. Chacko, D. G. Kanhere, and S. A. Blundell,
  Phys. Rev. B {\bf 71}, 155407 (2005).

\bibitem{Lee-small-sodium}
  M.-S. Lee, S. Chacko, and D. G. Kanhere,
  J. Chem. Phys. {\bf 123}, 164310 (2005).

\bibitem{Lee-58}
  M.-S. Lee and D. G. Kanhere,
  Phys. Rev. B {\bf75}, 125427 (2007).

\bibitem{Aguado}	
  A. Aguado and J. M. L{\'o}pez,
  Phys. Rev. Lett. {\bf 94}, 233401 (2005).

\bibitem{Aguado30}
  A. Aguado and J. M. L{\'o}pez,
  Phys. Rev. B {\bf 74}, 115403 (2006).

\bibitem{Jarrold-Ga}
  G. A. Breaux, D. A. Hillman, C. M. Neal, R. C. Benirschke, and M. F. Jarrold,
  J. Am. Chem. Soc. {\bf 126}, 8628 (2004).

\bibitem{Kavita}
  K. Joshi, S. Krishnamurty, and D. G. Kanhere,
  Phys. Rev. Lett. {\bf 96}, 135703 (2006).

\bibitem{Jarrold-Al}
  G. A. Breaux, C. M. Neal, B. Cao, and M. F. Jarrold,
  Phys. Rev. Lett. {\bf 94}, 173401 (2005).

\bibitem{Au19-20}
  S. Krishnamurty, G. Shafai, D. G. Kanhere, and M. J. Ford,
  Cond-mat/0612287 (to be published).

\bibitem{Rothlisberger}
  U. R{\"o}thlisberger and W. Andreoni,
  J. Chem. Phys. {\bf 94}, 8129 (1991)

\bibitem{MD}
  M. C. Payne, M. P. Teter, D. C. Allan, T. A. Arias, and J. D. Joannopoulos,
  Rev. Mod. Phys. {\bf 64}, 1054 (1992).

\bibitem{vanderbilt}
  D. Vanderbilt, 
  Phys. Rev. B {\bf 41}, 7892 (1990).

\bibitem{VASP}
  Vienna {\em ab initio } simulation package,
  Technische Universit\"at Wien (1999);
  G. Kresse and J. Furthm\"uller,
  Phys. Rev. B {\bf 54}, 11169 (1996).

\bibitem{Basin}
  Z. Li and H. A. Scheraga, Proc. Natl. Acad. Sci. {\bf84}, 6611 (1987); 
  D. J. Wales and J. P. K. Doye, J. Phys. Chem. A {\bf101}, 5111 (1997).

\bibitem{genetic1}
  M. Iwamatsu,
  J. Chem. Phys. {\bf112}, 10976 (2000)

\bibitem{genetic2}
  J. A. Niesse and H. R. Mayne,		
  J. Chem. Phys. {\bf105}, 4700 (1996)

\bibitem{ab initio}
  R. G. Parr and W. Yang, {\it The Density Functional Theory of Atoms and 
  Molecules} (Oxford university Press, New York, 1989). 

\bibitem{Mh1}
  A. M. Ferrenberg, R. H. Swendsen,
  Phys. Rev. Lett. {\bf 61}, 2635 (1988).

\bibitem{Mh2}
  P. Labastie and R. L. Whetten,
  Phys. Rev. Lett. {\bf 65}, 1567 (1990).
	
\bibitem{Abhijat}
  A. Vichare, D. G. Kanhere, and S. A. Blundell,
  Phys. Rev. B {\bf 64}, 045408 (2001).  

\end{thebibliography}
\end{document}